\documentclass[pdflatex,sn-mathphys-num]{sn-jnl}% Math and Physical Sciences 

\usepackage{graphicx}%
\usepackage{multirow}%
\usepackage{amsmath,amssymb,amsfonts}%
\usepackage{amsthm}%
\usepackage{mathrsfs}%
\usepackage[title]{appendix}%
\usepackage{xcolor}%
\usepackage{textcomp}%
\usepackage{manyfoot}%
\usepackage{booktabs}%
\usepackage{algorithm}%
\usepackage{algorithmicx}%
\usepackage{algpseudocode}%
\usepackage{listings}%
\theoremstyle{thmstyleone}%
\theoremstyle{thmstyletwo}%

\theoremstyle{thmstylethree}%

\begin{document}

\title[MASA: Improved Loc. by Beamformed Speech]{Multi-agent Auditory Scene Analysis: Improved Localization Speed and Robustness by Multi-beamformed Speech Quality Feedback}


\author[1]{\fnm{Caleb} \sur{Rascon}}\email{caleb@unam.mx}

\affil[1]{\orgdiv{Instituto de Investigaciones en Matematicas Aplicadas y en Sistemas}, \orgname{Universidad Nacional Autonoma de Mexico}, \orgaddress{\street{Circuito Escolar 3000}, \city{Coyoacan}, \postcode{03740}, \state{Ciudad de Mexico}, \country{Mexico}}}


\abstract{A real-time auditory scene analyzer (ASA) aims to carry out the tasks of locating, separating and classifying the sound sources present in a given acoustic environment. Recently, an effort has been made into modelling an ASA as a multi-agent system, with each one of its agents performing one of the aforementioned tasks and communicating their results to the rest of their peer agents. These communication routes are used as feedback loops to fix local errors at a global level, providing robustness while reducing local complexity. An example of the benefits of this approach is the optimization of speech quality by correcting in real-time the estimated location of the speech source of interest. However, their optimization speed has been shown to be considerably slow. One possible reason is that it solely relies on a series of single quality estimations (provided by a reference-free quality estimator model) that vary considerably from one window to the next, which results in a difficult search space to optimize. In this work, a new optimization mechanism is proposed that instead relies on a series of sets of quality estimations over a range of locations, providing a clearer view of the search space, simplifying its optimization. The proposed ASA now has a considerably smaller optimization time, is more accurate, and is more stable when being evaluated in real-life acoustic scenarios to correct higher levels of localization errors, all while being less complex than previous efforts. The only trade-off is that there is an increase in the response time of the quality estimation agent, but the complete ASA is still able to run in real-time. The performance shown in this work again shows the benefits of modelling an ASA as a multi-agent system.}


\maketitle

\section{Introduction}\label{sec:introduction}

A real-time auditory scene analyzer (ASA) aims to carry out, and in conjunction, the tasks of locating, separating, and classifying sound sources in a given acoustic environment \cite{bregman1994auditory,wang2006computational}. Doing so can provide important benefits to application areas such as bioacoustics \cite{stowell2017computational}, hearing aid research \cite{zhang2019study,green2022speech}, search and rescue scenarios \cite{nakadai2017development}, human-robot interaction \cite{rascon2015integration}, etc.

Usually, in an ASA, the aforementioned tasks are carried out in a linear manner \cite{nakadai2017hark,grondin2022odas}: all the sound sources are located; then, using their locations, each is separated into its own channel; and, once separated, information is extracted from each source. However, doing so makes the whole ASA sensitive to localization errors (the first task), which begets the development of more complex localization techniques that are more robust \cite{rascon2017localization}. This, in turn, increases the computational load and the global response time. Thus, carrying out these tasks in a linear manner results in local task-level errors having a global impact.

\begin{figure*}[h]
	\centering
	\includegraphics[width=0.9\textwidth]{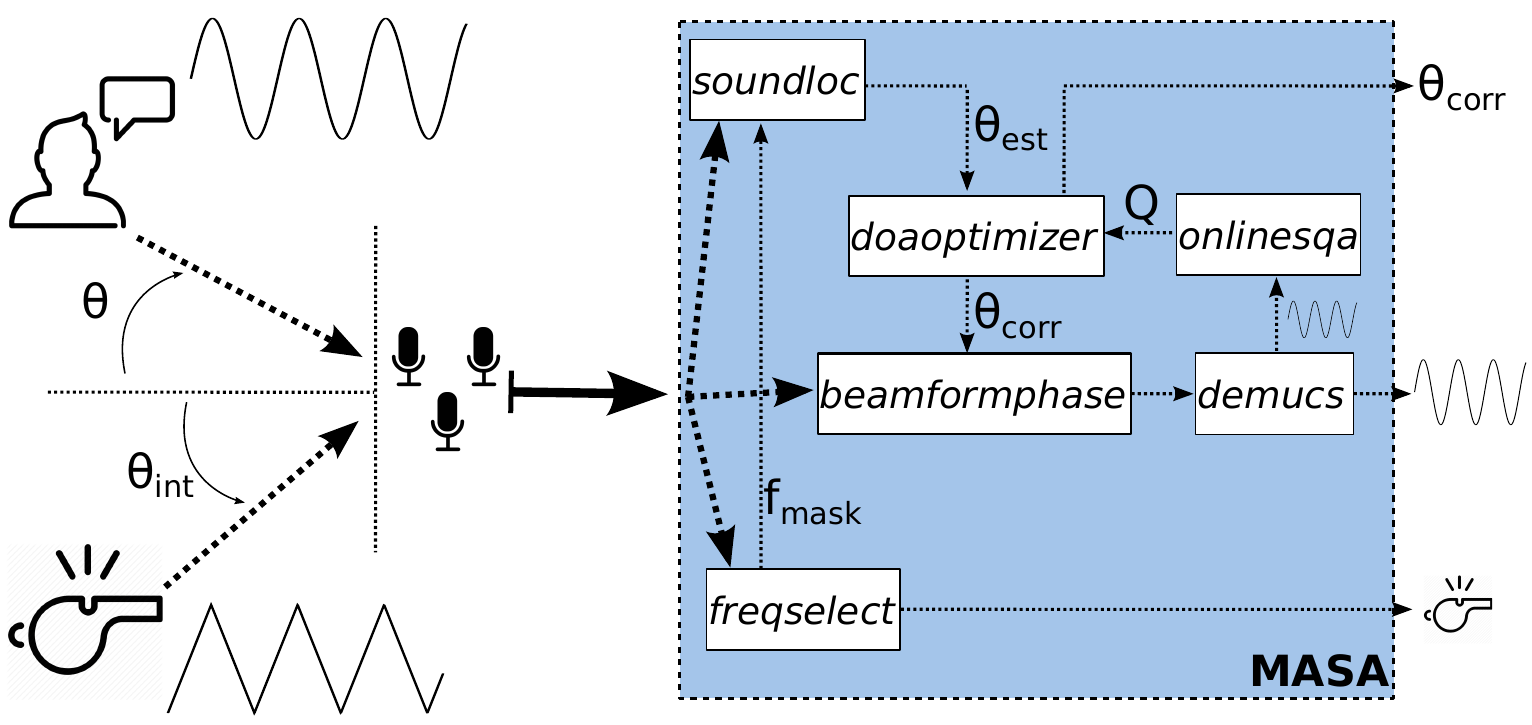}
	\caption{Current version of the multi-agent auditory scene analyzer presented in \cite{rascon2026multi}.}\label{fig:masa}
\end{figure*}

To counter this issue, recently, there has been an effort \cite{rascon2025direction,rascon2026multi} to model an ASA as a multi-agent system, where each agent is in charge of a particular task (localization, separation or classification) and sends its results to other agents as a type of feedback. This inter-agent interaction has been shown to provide local error robustness, while reducing overall global complexity. The current version of the multi-agent auditory scene analyzer (MASA) \cite{rascon2026multi} is shown in Figure \ref{fig:masa}.

As it can be seen, there are several communication loops between agents. The focus of this work is the feedback loop between the agents in charge of the separation tasks (named \textit{beamformphase} and \textit{demucs}) and the result of the localization task ($\theta_{est}$ being outputted by the \textit{soundloc} agent). There are two other agents that are part of this loop: the quality estimation agent (\textit{onlinesqa}) and the optimization agent (\textit{doaoptimizer}). The aim of this feedback loop is to find the location ($\theta_{corr}$) that optimizes the separation quality ($Q$). It has shown to be able to globally correct localization errors at the local agent-level in real-time \cite{rascon2025direction}, reducing the need of highly complex localization techniques \cite{rascon2017localization}. Later, the work in \cite{rascon2026multi} improves the optimization process with which the localization error is corrected by introducing `memory' of the best location it has found. It provides much more robust results, with varying degrees of success in different acoustic scenarios \cite{rascon2026multi}.

In both works \cite{rascon2025direction,rascon2026multi}, the optimization process relies on a series of single quality estimations provided by a reference-free model called Squim \cite{kumar2023torchaudio}, while an Adam-based optimizer \cite{kingma2014adam} is employed to carry out the optimization process. However, in both works, the optimization mechanism requires a long time to reach the correct location, while providing high variability (low stability) in its location correction. Additionally, the range of localization errors that both mechanisms can reliably correct is relatively small, while being moderately complex.

In this work, a new optimization mechanism is proposed that relies, instead, on a series of sets of quality estimations at each optimization iteration. Each quality estimation inside each set is `steered' towards different locations, providing a clearer view of the search space. This, in turn, simplifies the optimization process to a simple weighted average, using the quality estimation values as weights. From these two simple operators being carried out in both the quality estimation agent and the optimization agent, an optimization process emerges that not only provides much more robust and stable results than the previous optimization mechanisms, but is able to correct a wider range of localization errors, with smaller optimization times, all while being less complex. This new optimization mechanism is further evidence of the benefits of modeling an ASA as a multi-agent system where, by dealing with local errors globally, robust results are obtained with less overall complexity.

This work is structured as follows: Section \ref{sec:background} provides background on modeling an auditory scene analyzer as a multi-agent system, as well as details of the previous optimization mechanisms for correcting localization errors through quality feedback; Section \ref{sec:newoptim} describes the proposed optimization mechanism; Section \ref{sec:results} details the evaluation material and methodology, presents the results of the parameter characterization, and provides a comparison with the previous optimization mechanisms; and Section \ref{sec:conclusion} presents the concluding remarks as well as future work.

\section{Background of Multi-agent Auditory Scene Analysis} \label{sec:background}

\begin{figure*}[h]
	\centering
	\includegraphics[width=0.9\textwidth]{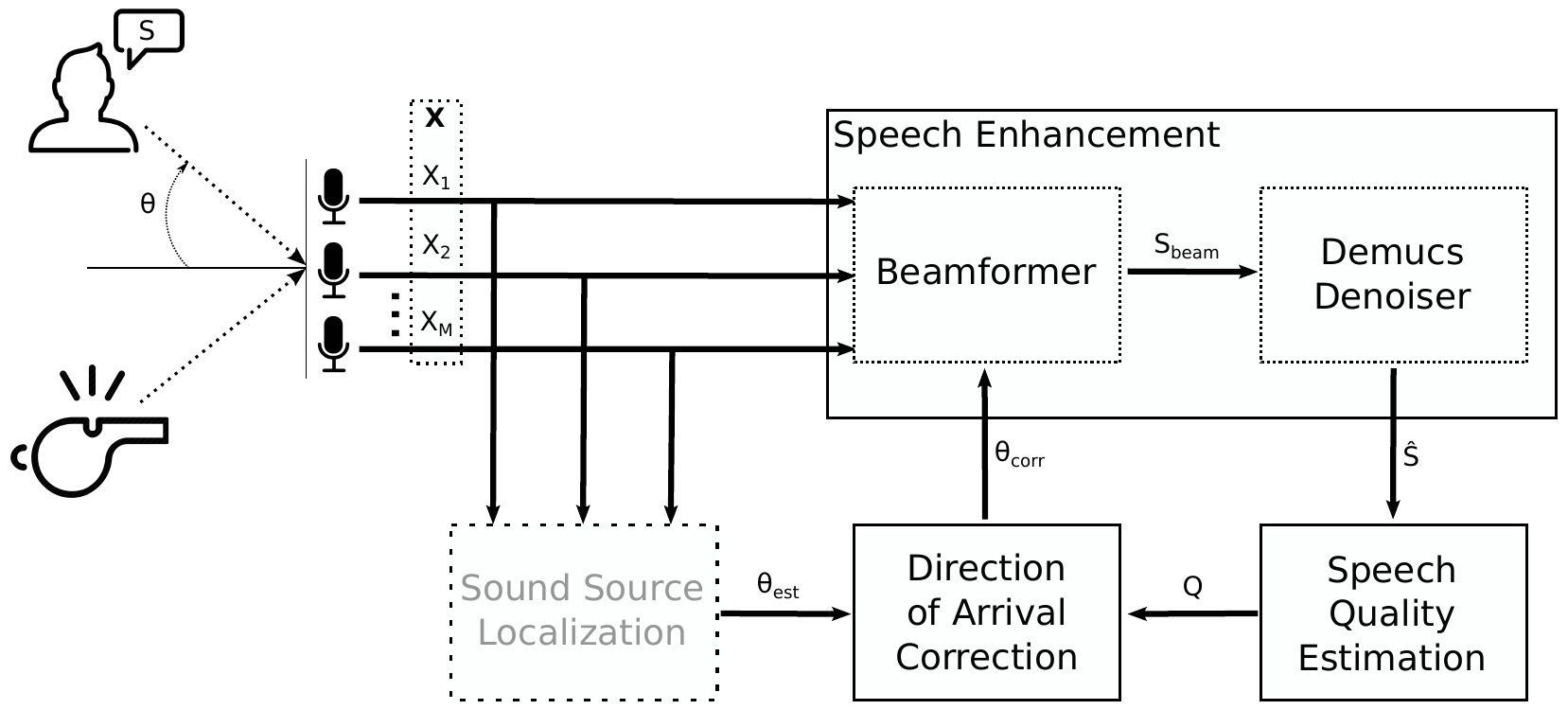}
	\caption{Location (direction of arrival) correction as presented in  \cite{rascon2025direction}.}\label{fig:doacorrection}
\end{figure*}

Recently, carrying out the tasks of locating, separating and classifying the sound sources in the environment in conjunction has been modeled as a multi-agent system \cite{rascon2025direction,rascon2026multi}, with each agent carrying out a different task in a non-linear manner. This approach has shown to be globally robust against agent-level errors, with an initial study correcting localization errors through the feedback of the separated speech quality \cite{rascon2025direction}. The target speech is separated by using its location through a pre-beamforming stage; thus, if the location is incorrect, the speech quality degrades. The main idea is to correct in real-time this location by optimizing the speech quality, as shown in Figure \ref{fig:doacorrection}.

\clearpage

As it can be seen, the speech enhancement module is conformed by a beamformer that is steered towards the speech source of interest, and its output is fed to a Demucs-based speech enhancer \cite{defossez20_interspeech}.

To estimate the speech quality without the need of a reference signal, Squim \cite{kumar2023torchaudio} was employed by feeding it a series of a 3-second windows ($t_w$). From each window, Squim estimates the signal-to-distortion ratio (SDR) as a means to estimate the amount of interference still present in the output. It was observed that these estimations vary considerably from window to another, thus, exponential smoothing was employed as shown in Equation \ref{eq:expsmooth}.

\begin{equation}
\label{eq:expsmooth}
Q_k \gets \alpha Q_k + (1-\alpha)Q_{k-1}
\end{equation}

where $Q_k$ is the quality estimation at the $k$ moment in time (ie. optimization iteration index), and $\alpha$ is a smoothing factor in the range of $[0,1]$ with higher values providing smoother results but less responsiveness to underlying changes, and vice-versa. It was found that an $\alpha$ value of 0.9 was a good balance between providing a smooth output and still being somewhat responsive to underlying changes.

These smoothened results were optimized using an Adam-based optimization mechanism \cite{kingma2014adam}, where it dynamically changed the updating factor of the controlled variable (which in this case is the sound source location\footnote{In this work, the location of a sound source is equated to a direction of arrival.}, $\theta_c$) during its optimization process, considering the gradient of the optimized value (which in this case is the quality estimation, $Q$). For completeness sake, this optimization mechanism is presented in Algorithm \ref{alg:optimization1}.

\clearpage

\begin{algorithm}[h]
	\caption{Optimization mechanism from \cite{rascon2025direction}.}\label{alg:optimization1}
	\begin{algorithmic}[1]
		\Require $t_h$: time hop; $\eta$: learning rate
		\Require $\beta_m$: mean update factor ($0.9$); $\beta_v$: variance update factor ($0.999$)
		\State Initialize: $\nabla m \gets 0$
		\State Initialize: $\nabla v \gets 0$
		\State Initialize: $\theta_p \gets 0$ \Comment{force an appropriate gradient at the start}
		\State Initialize: $\theta_c \gets \theta_{est}$ \Comment{from sound localization agent}
		\State Initialize: $Q_p \gets 0$
		\State Initialize: $Q_c \gets 0$
		\Loop
		\State $\theta_c \to \theta_{corr}$  \Comment{to speech enhancement agent}
		\State \textbf{wait} $t_h$ seconds
		\State $Q_p \gets Q_c$
		\State $Q_c \gets Q$  \Comment{from quality estimation agent}
		\State $Q_c \gets 100 - Q_c$ \Comment{convert $Q_c$ to a minimizable value}
		\State $\nabla Q \gets \frac{Q_c-Q_p}{\theta_c-\theta_p + \epsilon}$ \Comment{calculate quality gradient}
		\State $\nabla m \gets \beta_m \nabla m + (1 - \beta_m)\nabla Q$ \Comment{update momentum and variance}
		\State $\nabla v \gets \beta_v \nabla v + (1 - \beta_v)\nabla Q^2$
		\State $\theta_p \gets \theta_c$
		\State $\theta_c \gets \theta_c - \eta \frac{\nabla m}{\sqrt{\nabla v} + \epsilon}$ \Comment{update $\theta_c$ with both gradients}
		\EndLoop
	\end{algorithmic}
\end{algorithm}

The Adam optimization method requires, however, that the objective function be twice continuously differentiable. Although it was found that the smoothened quality estimations provided an objective function that was `close' to satisfying this requirement, it was also found that it often overshoots the correct location, which takes a considerable amount of time to correct. To this effect, a new optimization mechanism was presented in \cite{rascon2026multi} that `remembers` the best location ($\theta_{best}$) which has provided the highest quality ($Q_{best}$). If a certain amount of optimization iterations have gone by without a better quality estimation, the optimization process `resets' itself and re-starts from that known local optimum. Again, for completeness sake, this optimization mechanism is detailed in Algorithm \ref{alg:optimization2}, where all variables are initialized at $0$, unless stated otherwise.

\clearpage

\begin{algorithm}[h]
	\caption{Optimization mechanism from \cite{rascon2026multi}.}\label{alg:optimization2}
	\begin{algorithmic}[1]
		\Require $t_h$: time hop; $\eta$: learning rate
		\Require $\beta_m$: mean update factor ($0.9$); $\beta_v$: variance update factor ($0.999$)
		\State Initialize: $\theta_{hist} \gets [0,0,\dots,0]$; $i_{ncm} \gets 10$; $\theta_{best} \gets \mathit{None}$; $Q_{best} \gets Q_{MAX}$
		\State Initialize: $\theta_c \gets \theta_{est}$ \Comment{from sound localization agent}
		\Loop
		\State \textbf{shift} $\theta_{hist}$ one position
		\State $\theta_{hist}[0] \gets \theta_c$
		\State $i_{calc} \gets i_{calc}+1$
		\If{$i_{calc} \geq \mathit{length}(\theta_{hist})$}
		\If{$Q_c < Q_{best}$}
		\State $Q_{best} \gets Q_c$
		\State $\theta_{best} \gets \theta_c$
		\Else
		\State $i_{nocorr} \gets i_{nocorr}+1$ 
		\If{$i_{nocorr} \geq i_{ncm}$}
		\State $\theta_c \gets \theta_{best}$ \Comment{reset to best location}
		\State $\theta_p \gets 0$
		\State $\theta_{hist} \gets [0,0,\dots,0]$
		\State $\theta_{best} \gets \mathit{None}$
		\State $Q_p \gets 0$; $Q_c \gets 0$
		\State $i_{nocorr} \gets 0$; $i_{calc} \gets 0$
		\EndIf
		\EndIf
		\EndIf
		\State $\theta_c \to \theta_{corr}$ \Comment{to speech enhancement agent}
		\State \textbf{wait} $t_h$ seconds
		\State $Q_p \gets Q_c$
		\State $Q_c \gets Q$ \Comment{from quality estimation agent}
		\State $\nabla Q \gets \frac{Q_c-Q_p}{\theta_c-\theta_p + \epsilon}$ 
		\State $\nabla m \gets \beta_m \nabla m + (1 - \beta_m)\nabla Q$ 
		\State $\nabla v \gets \beta_v \nabla v + (1 - \beta_v)\nabla Q^2$
		\State $\theta_p \gets \theta_c$
		\State $\theta_c \gets \theta_c - \eta \frac{\nabla m}{\sqrt{\nabla v} + \epsilon}$
		\EndLoop
	\end{algorithmic}
\end{algorithm}

Algorithm \ref{alg:optimization2} has two omissions from the implementation shown in \cite{rascon2026multi}: the results of the localization agent are merged with the ones corrected by the optimization process (which is not shown); and the quality estimation is normalized so that the optimization process minimizes it, which is expected by the Adam technique (which is also not shown). Both omissions can be ignored for the topic of this work without loss of generalization.

The variation of the optimization mechanism presented in \cite{rascon2026multi} reduces localization errors, provides more consistent results, and increases the rate of successful runs, compared to the optimization mechanism presented earlier in \cite{rascon2025direction}. However, it does not provide any reduction in optimization time and is more complex (although it is still able to run in real-time).

It is essential to note that, even though Squim's response time makes it viable to be run in real-time, it was not built to be run in a causal manner: it assumes that it is being fed a complete audio recording. In the real-time scenario presented in this work, a series of audio windows are fed instead. To this effect, it has been shown \cite{rascon2025direction} that Squim estimates a quality from each audio window that is different from the rest of the windows of the series, even when the true quality of all windows is the same. This results in a series of single quality estimations that vary considerably through time, creating a very difficult search space to optimize for an Adam-based technique. Thus, low update factors end up being required to find the correct location, but do so in a very slow manner. In this work, it is hypothesized that a clearer view of the search space for each window (ie. optimization iteration) can simplify the optimization process, as it is discussed in the following section.

\section{Proposed Optimization Mechanism} \label{sec:newoptim}

Instead of relying on a series of single quality estimations that vary through time, in this work, a new optimization mechanism is proposed where a series of sets of quality estimations are carried out based on the outputs of a set of beamformed locations. To this effect, the quality is not only estimated from the requested location ($\theta_{mid}$), but also from a number of locations ($\theta_{steps}$) inside a given location range ($\theta_{range}$). Since the same audio window is used for all quality estimations (just ``steered'' towards different locations), a clearer view is provided of the search space at each optimization iteration. From here on forward, for simplicity, estimating the quality at a given location is referred to as ``testing a location''.

From preliminary evaluations, it was observed that using the output of the whole speech enhancement module (the output of the Demucs-based speech enhancer) resulted in a flat search space, while requiring a considerable amount of computing resources. Thus, it was decided to use only the output of the beamformer stage of the speech enhancement module, which resulted in a more convex search space as well requiring less computing resources. Furthermore, in addition to the phase-based frequency masking (PHASE) beamformer \cite{rascon2021corpus} used in the original implementation of the speech enhancer, the minimum variance distortion-less response (MVDR) beamformer \cite{levin1964maximum, capon1969high,habets2009new} was also evaluated since it is less prone to output phase-induced artifacts with real-life reverberant acoustic scenarios (further details on the evaluation methodology are provided in Section \ref{sec:results}).

Once the set of quality estimations ($Q_{set}$) are calculated for all the locations ($\theta_{set}$) in the given $\theta_{range}$, these are unified into one quality estimation ($\theta_c$) via a simple weighted average, with the quality estimation values ($Q_{set}$) acting as weights for their corresponding locations ($\theta_{set}$). Algorithm \ref{alg:newquality} presents how the set of quality estimations are calculated, while the new proposed optimization mechanism is presented in Algorithm \ref{alg:newoptimization}.

\clearpage

\begin{algorithm}[h]
	\caption{Quality estimation based on beamform search.}\label{alg:newquality}
	\begin{algorithmic}[1]
		\Require $t_w$: window length
		\State Initialize: $\theta_{set}$, $Q_{set}$
		\State \textbf{wait} for quality request $\gets$ ($\theta_{mid}$, $\theta_{range}$, $\theta_{steps}$) \Comment{from optimization agent}
		\State $x \gets \mathit{get\_audio\_from\_mics}(t_w)$
		\For{\textbf{each} $\theta \in \mathit{linspace}([\theta_{mid}-\theta_{range},\theta_{mid}+\theta_{range}], \mathit{steps}:\theta_{steps})$}
		\State $w \gets \mathit{calculate\_weights}(\theta)$
		\State $y \gets \mathit{beamform}(x,w)$
		\State $q \gets \mathit{Squim}(y)$ \Comment{estimating quality at  $\theta$}
		\State $\theta_{set}.\mathit{append}(\theta)$
		\State $Q_{set}.\mathit{append}(q)$
		\EndFor
		\State \textbf{respond} to quality request $\to$ ($\theta_{set}$, $Q_{set}$) \Comment{to optimization agent}
		\State reset($\theta_{set}$, $Q_{set}$)
	\end{algorithmic}
\end{algorithm}

\begin{algorithm}[h]
	\caption{Proposed optimization mechanism.}\label{alg:newoptimization}
	\begin{algorithmic}[1]
		\Require $t_h$: time hop
		\Require $\theta_{range}$: location range below and above $\theta_{mid}$
		\Require $\theta_{steps}$: steps to calculate inside $\theta_{range}$
		\State Initialize: $\theta_{mid} \gets \theta_{est}$ \Comment{from sound localization agent}
		\Loop
		\State \textbf{wait} for $t_h$ seconds
		\State \textbf{request} quality estimation ($\theta_{mid}$, $\theta_{range}$, $\theta_{steps}$) \Comment{to quality estimation agent}
		\State ($\theta_{set}$, $Q_{set}$) $\gets$ \textbf{response} from quality estimation agent
		\State $\theta_c \gets \frac{\sum_{i}\theta_{set}[i] * Q_{set}[i]}{\sum_{i}Q_{set}[i]}$
		\If{$\theta_c > \theta_{set}[\mathit{end}]$}
		\State $\theta_c \gets \theta_{set}[\mathit{end}]$
		\ElsIf{$\theta_c < \theta_{set}[0]$}
		\State $\theta_c \gets \theta_{set}[0]$
		\EndIf
		\State $\theta_c \to \theta_{corr}$ \Comment{to speech enhancement agent}
		\State $\theta_{mid} \gets \theta_{c}$
		\EndLoop
	\end{algorithmic}
\end{algorithm}

As it can be seen, the optimization process is considerably simplified. However, it is important to mention that the quality estimation response time is extended by the amount of tested locations. Fortunately, as reported in \cite{rascon2025direction}, with moderate hardware, Squim's response time is between $0.0538$ s. and $0.0704$ s. for $t_w=3.0$, which means that there is enough time to test several locations for a given $t_w$ window while still remaining below the real-time threshold.

\begin{figure*}[h]
	\centering
	\includegraphics[width=0.9\textwidth]{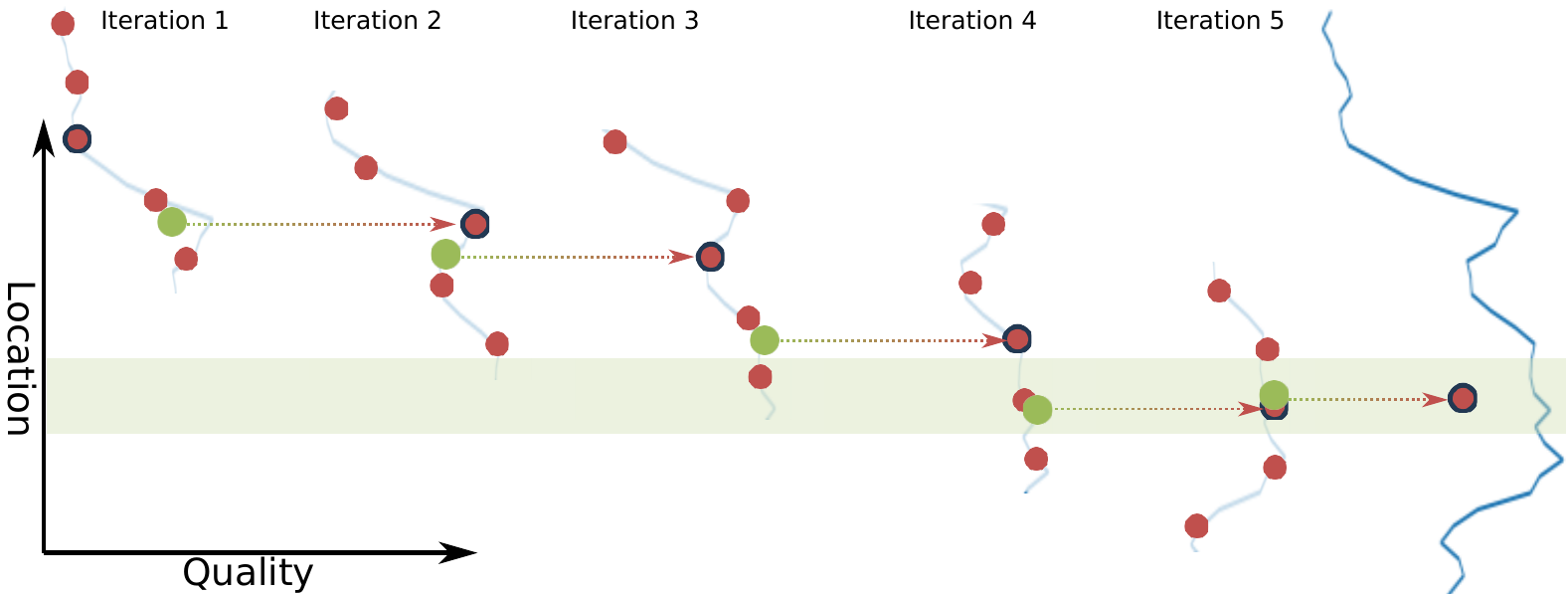}
	\caption{A diagram summarizing the reasoning behind the proposed optimization mechanism.}\label{fig:optimization}
\end{figure*}

The reasoning behind this new optimization mechanism is that carrying out a weighted average, as presented in Algorithm \ref{alg:newoptimization}, can be considered as a simplified interpolation of the quality estimated in a range of locations ($\theta_{range}$). It should provide a clearer view of the current search space compared to only having the quality estimated from one location. By forcing the weighted average to be inside the location range, the next optimization iteration will estimate the quality of a location range overlapping with the current one, which reduces overshooting. This reasoning is summarized in Figure \ref{fig:optimization}. The solid blue line in the right is the complete location-vs-quality search space, which is unknown to the optimization process. Each blue line to its left represents one optimization iteration, along with its corresponding search space range (also unknown) which is limited by its respective location range. The red dots with black border are the $\theta_{mid}$ of their respective optimization iteration; all red dots represent a tested location; green dots are the weighted average of their respective iteration. The green area is where the correct location resides. As it can be seen, by testing a set of locations at each optimization iteration (instead of just one), and carrying out a simple weighted average, the whole optimization process reaches the correct location. 

It is important to note that the quality estimation agent is only providing a set of quality estimations and that the optimization agent is only carrying out a weighted average of these estimations. The optimization process is not being carried out in a single agent, as was the case in the previous optimization mechanisms. Meaning, the optimization process is actually a behavior that emerges from the interaction between these two agents. Thus, not only is the optimization process greatly simplified, but thanks to these agents' interaction, a robust optimization process emerges in a global manner. This is even more evidence of the benefits of modeling the auditory scene analysis as a multi-agent system: robust results with less overall complexity.

Furthermore, the proposed optimization mechanism can be `calibrated' with both the $\theta_{range}$ and $\theta_{steps}$ parameters, which act in conjunction as a type of `update factor' for the emerging optimization mechanism. However, it is essential to characterize the behavior of the proposed optimization mechanism with different values of the $[\theta_{range},\theta_{steps}]$ pair, referred here on forward as an `optimization configuration'.

\section{Evaluation and Results}\label{sec:results}

In this section, several evaluations are carried out to characterize both the parameters of the proposed optimization mechanism: the tested location range ($\theta_{range}$) and the amount of tested locations inside such range ($\theta_{steps}$). This characterization has the objective to establish a range of values for each parameter that provide a small optimization time as well as a localization correction that is robust against varying acoustic scenarios and number of interferences. Additionally, a comparison is provided between the proposed optimization mechanism (using the recommended values of the aforementioned parameters) and the previous optimization mechanism presented in \cite{rascon2026multi}.

\subsection{Material}

The Acoustic Interactions for Robot Audition (AIRA) corpus \cite{rascon2018acoustic} is conformed by a considerable amount of recordings, with varying number of speech sources, in different real-life acoustic settings, with different microphone array geometries. In both \cite{rascon2025direction} and \cite{rascon2026multi}, the ``Anechoic Chamber'' scenario was used to evaluate their respective optimization mechanisms, using a 3-microphone array in the presence of 2 speech sources. In \cite{rascon2026multi}, additionally, the ``Office A'' scenario, as well as the ``Department Store'' scenario, (both using a similar 3-microphone array) were also employed to evaluate the generality of its optimization mechanism with varying amounts of speech sources.

The ``Department Store'' scenario is comparable to a Walmart or Tesco, and has a high overall noise level (63 dB SPL) and a moderate level of reverberation ($\tau_{60}=0.16s$). The ``Office A'' scenario is comparable to a computer laboratory, and has a lower overall noise level (52 dB SPL) and a higher level of reverberation ($\tau_{60}=0.20s$).

However, in both \cite{rascon2025direction} and \cite{rascon2026multi}, it was only with the ``Anechoic Chamber'' scenario that parameters were characterized, which has virtually no noise nor reverberation. Additionally, when employing the ``Office A'' and ``Department Store'', only a small amount of localization error was explored to be corrected ($10^\circ$).

In this work, the ``Office A'' scenario, using a 3-microphone array in the presence of 2 speech sources, is employed to characterize the parameters in a real-life environment. Afterwards, the proposed optimization mechanism is evaluated with varying amounts of speech sources in that scenario, as well as in the ``Department Store'' scenario of the AIRA corpus.

The 3-microphone array geometry is an equilateral triangle with an inter-microphone distance of $0.18$ m. The 2 speech sources are located in $0^\circ$ and $90^\circ$ respectively, 1 m. away from the center of the microphone array. Unless stated otherwise, the speech source of interest is the one located at $0^\circ$.

\subsection{Methodology}

To carry out repeatable evaluations, an agent feeds the multi-channel recording in real-time to the rest of the agents to emulate a live microphone using the JACK Audio Connection Kit \cite{JACKAudio}. This emulation agent reads 1024-sample windows at $48$ kHz (equating to around $0.021$ s. latency), as it was captured for the AIRA corpus. The emulation agent also provides a $16$ kHz stream of the same multi-channel recording, as it is required by Squim. Finally, to provide sufficient evaluation time, the recording (which is originally $30$ s. long) is repeated $4$ times, resulting in an evaluation time of approximately $120$ s. Additionally, an artificial localization error ($\theta_{error}$) is applied at the beginning of the run, which the proposed optimization mechanism is expected to correct.

The quality estimation agent receives the multi-channel audio data directly from the emulation agent, storing the last $t_w=3.0$ s. of audio information and awaiting requests from the optimization agent. In turn, the optimization agent waits $t_h=0.25$ s. to carry out a request. It is important to mention that $t_h$ could be lowered to reduce the overall response time, but it was set at this value to reduce computational load.

The computer used to carry out all of the evaluations uses an AMD Ryzen 5 3600 (6 cores with a 3.6 GHz base clock) and a GTX 1050 Ti with 4 GB of VRAM. This setup was chosen because it is representative of a computer with moderate computational capabilities that is within range of the current average consumer.

\subsection{Variables and Metrics}

Two beamforming techniques are evaluated: the phase-based frequency masking beamformer (PHASE) \cite{rascon2021corpus} used in the original implementation of the speech enhancer, and the popular minimum variance distortion-less response beamformer (MVDR) \cite{levin1964maximum, capon1969high,habets2009new}.

Additionally, Squim not only provides an estimation of the signal-to-distortion ration (SDR), but also of the short-time objective intelligibility (STOI) measure \cite{taal2010short} as well as of the perceptual evaluation of speech quality (PESQ) measure \cite{recommendation2001perceptual}. All three evaluation measures are evaluated.

Finally, as for optimization configurations, defined by the $\theta_{range}$ and $\theta_{steps}$ parameters, a range of values for each are evaluated. For $\theta_{range}$, the values are $[5.0, 10.0, 15.0, 20.0, 25.0]$; these values represent the tested location range (in degrees) that the quality estimation agent is going to use. For $\theta_{steps}$, the values are $[3, 5, 7, 9, 11]$; these values represent the amount of locations that are going to be tested by the quality estimation agent along $\theta_{range}$, including the requested `middle' location ($\theta_{mid}$).

To this effect, there are a total of four evaluation variables: beamform type, quality type, $\theta_{range}$, and $\theta_{steps}$. For each combination of values of these evaluation variables, $10$ `runs' are carried out running the real-time emulated $120$ s. recording, while the series of corrected locations provided by the optimization agent ($\theta_{corr}$) is recorded for each run.

Unless stated otherwise, each of the runs employed $\theta_{error}=15.0^\circ$, which is higher than when evaluating the optimization mechanism presented in \cite{rascon2026multi} in different acoustic settings.

Several metrics are used to characterize and evaluate the proposed optimization mechanism:

\begin{itemize}
	\item Accuracy (RMS): the root mean square localization error is used to measure how accurate is the optimization process. Only the last $3/4$ of the run is employed for calculating this metric so as to measure the accuracy where the optimization process should have stabilized. Additionally, by calculating this metric in this manner, it is also useful for discarding slow converging configurations, since these will have large errors at the start of the run. If the metric would had been calculated with the whole run, it would had suffered from a large bias towards $\theta_{error}$ in these cases.
	\item Variance ($\sigma_{RMS}$): the standard deviation of the localization error of the last $3/4$ of the run is used to measure the magnitude of the oscilations of the optimization process. It is representative of the stability of the system.
	\item Optimization time ($t_{opt}$): the time it takes to reach the $[-5^\circ,5^\circ]$ range around the correct location of the target speech source is used to measure the speed of the optimization process. For simplicity, this range is referred to as the $[\theta_{correct}]$ range from here on forward.
	\item Quality estimation latency ($t_{lat}$): the response time of the quality estimation agent is used to assess the overall quality estimation latency. It is measured from the moment the optimization agent makes a request to the quality estimation agent, to the moment the optimization agent receives a response from the quality estimation agent.
\end{itemize}

Additionally, a `good run' is defined as one that has a RMS below $5^\circ$, while the `good run' percentage is measured as the amount of `good runs' divided the amount of runs for each evaluation variable combination. The `good run' percentage metric is used to discard inconsistent optimization configurations that perform well only some of the time.

\subsection{Quality Type and Beamforming Technique}

The mean accuracy of all the runs using each quality type are shown in Table \ref{tab:qualitytype}. As it can be seen, the STOI quality type provided the best accuracy by a large margin. Since these results considered all of the evaluation variable combinations, even the least accurate optimization configurations when using STOI tend to be more accurate than when using SDR or PESQ. Therefore, the STOI quality type is used from here on forward.

\begin{table}[h]
	\caption{Mean accuracy for each quality type.}\label{tab:qualitytype}
	\begin{tabular*}{13.5em}{cc}
		\toprule%
		Quality Type & RMS ($^\circ$) $\downarrow$ \\
		\midrule
		PESQ  & $8.0633$\\
		SDR  & $13.7106$\\
		\textbf{STOI}  & $\mathbf{2.0247}$\\
		\botrule
	\end{tabular*}
\end{table}

The overall `good run' percentages when using each beamforming technique (PHASE and MVDR), along with the STOI quality type, are shown in Table \ref{tab:beamformtype}. As it can be seen, MVDR outperforms PHASE considerably. Again, these results considered all evaluation variable combinations (that use STOI), which means that the least performant optimization configurations when using MVDR tend to perform better than when using PHASE. Therefore, MVDR is used from here on forward.

\begin{table}[h]
	\caption{Overall `good run' percentage for each beamforming technique, using STOI.}\label{tab:beamformtype}
	\begin{tabular*}{27em}{cc}
		\toprule%
		Beamforming Technique & Overall `good run' percentage $\uparrow$ \\
		\midrule
		PHASE  & $0.3187$\\
		\textbf{MVDR}  & $\mathbf{0.6244}$\\
		\botrule
	\end{tabular*}
\end{table}

\subsection{Characterization of Optimization Configurations}

\begin{figure*}[h]
	\centering
	\includegraphics[width=0.5\textwidth]{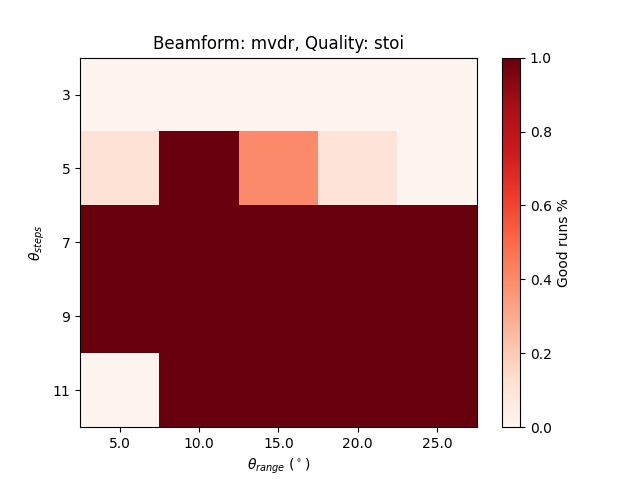}
	\caption{Optimization configurations vs. `good runs' percentage.}\label{fig:heatruns}
\end{figure*}

In Figure \ref{fig:heatruns}, a heatmap is shown of which $[\theta_{range},\theta_{steps}]$ optimization configurations provide the highest `good runs' percentage. As it can be seen, most configurations that satisfy $[\theta_{range}\geq 10^\circ,\theta_{steps}\geq 3]$ provide consistently high results.

\begin{figure*}[h]
	\centering
	\includegraphics[width=0.5\textwidth]{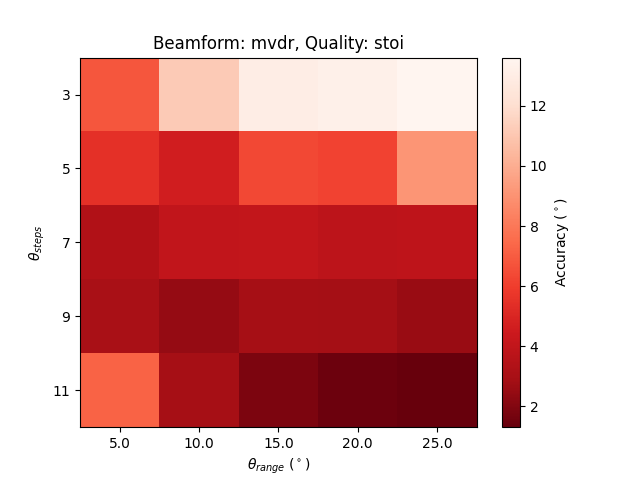}
	\caption{Optimization configurations vs. accuracy.}\label{fig:heatrms}
\end{figure*}

Additionally, in Figure \ref{fig:heatrms} a heatmap is shown of which $[\theta_{range},\theta_{steps}]$ configurations are the most accurate (measured by RMS). As it can be seen, in both heatmaps, there is a correlation between accuracy and high values of both $\theta_{range}$ and $\theta_{steps}$.

\clearpage

\begin{figure*}[h]
	\centering
	\includegraphics[width=0.5\textwidth]{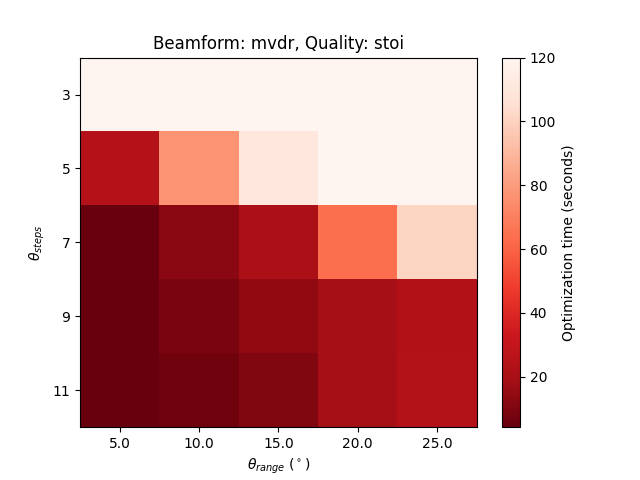}
	\caption{Optimization configurations vs. optimization time.}\label{fig:heatopttime}
\end{figure*}

However, in Figure \ref{fig:heatopttime}, a heatmap is shown of which $[\theta_{range},\theta_{steps}]$ configurations are the fastest (measured by $t_{opt}$). As it can be seen, there is a correlation between optimization time and low $\theta_{range}$ values with high $\theta_{steps}$ values, which is a partial opposite of the correlation observed in Figures \ref{fig:heatruns} and \ref{fig:heatrms}.

These results show that there are conflicting tendencies between accuracy and optimization speed, which is not unexpected given that a fast optimization usually results in unstable behavior \cite{ahn2022understanding}. Thus, a balance between accuracy and optimization speed is of interest. To this effect, a set of restrictions were applied to obtain a set of $[\theta_{range},\theta_{steps}]$ configurations that provided such balance. The restrictions are as follows:

\begin{itemize}
	\item For consistency: `good run' percentage of no lower than $1.0$.
	\item For accuracy: RMS below $3.0^\circ$.
	\item For speed: optimization time less than $15.0$ s..
\end{itemize}

The only optimization configurations that satisfied these restrictions are shown in Table \ref{tab:reccomb}.

\begin{table}[h]
	\caption{Recommended optimization configurations.}\label{tab:reccomb}
	\begin{tabular*}{27em}{cc|cccc}
		\toprule%
		$\theta_{range}$ & $\theta_{steps}$ & RMS $\downarrow$ & $\sigma_{RMS}$ $\downarrow$ & $t_{opt}$ $\downarrow$ & $t_{lat}$ $\downarrow$ \\
		\midrule
		\textit{20.0$^\circ$}  & \textit{5} & \textit{2.49$^\circ$} & \textit{0.36$^\circ$} &  \textit{8.73 s.} & \textbf{\textit{0.42 s.}} \\
		20.0$^\circ$  & 7 & 2.90$^\circ$ & 0.31$^\circ$ & 14.45 s. & 0.57 s. \\
		25.0$^\circ$  & 5 & 2.93$^\circ$ & 0.49$^\circ$ &  \textbf{6.46 s.} & \textbf{0.42 s.} \\
		25.0$^\circ$  & 7 & \textbf{1.86$^\circ$} & \textbf{0.23$^\circ$} & 10.14 s. & 0.57 s. \\
		\botrule
	\end{tabular*}
\end{table}

\begin{figure*}[h]
	\centering
	\includegraphics[width=0.5\textwidth]{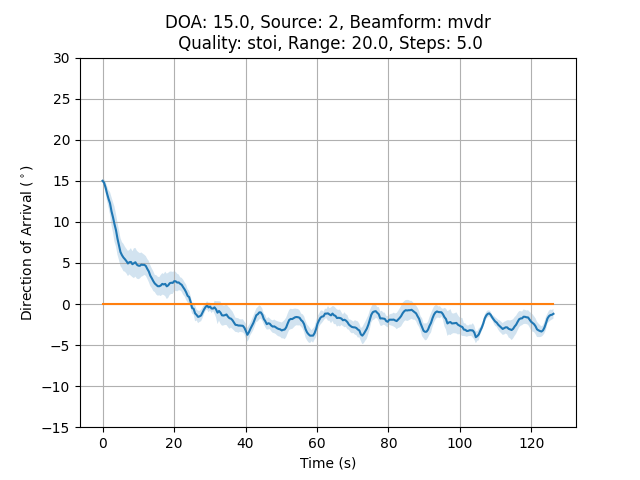}
	\caption{Behaviour of balanced optimization configuration $[\theta_{range}=20.0,\theta_{steps}=5]$.}\label{fig:optimrecom}
\end{figure*}

Although all the configurations in Table \ref{tab:reccomb} provide good results (the restrictions saw to that), they do so with different `priorities'. The $[\theta_{range}=25.0,\theta_{steps}=7]$ configuration provides the most accurate (lowest RMS) and most stable results (lowest $\sigma_{RMS}$), however, it provides one of the highest quality estimation latencies ($t_{lat}$) and the second highest optimization time ($t_{opt}$). On the other hand, the $[\theta_{range}=25.0,\theta_{steps}=5]$ configuration provides the lowest optimization time with one of the lowest quality estimation latencies, but at the cost of being the least accurate (highest RMS) with the least stable results (highest $\sigma_{RMS}$). The more balanced option is the $[\theta_{range}=20.0,\theta_{steps}=5]$ configuration, which provides a medium level of accuracy and results stability, with the second lowest optimization time and one of the lowest quality estimation latencies. In Figure \ref{fig:optimrecom}, the behavior of this optimization configuration is shown, where the dark blue line is the mean corrected location through time, and the area behind it is its variance.

\begin{figure*}[h]
	\centering
	\includegraphics[width=0.5\textwidth]{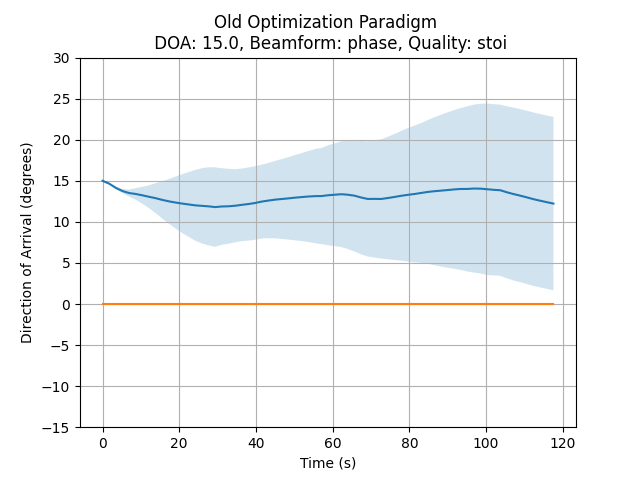}
	\caption{Behavior of the optimization mechanism presented in \cite{rascon2026multi}.}\label{fig:optimorig}
\end{figure*}

For comparison, Figure \ref{fig:optimorig} shows the behavior of the optimization mechanism in \cite{rascon2026multi} in the same scenario. It is important to remember that this optimization mechanism was originally evaluated in a similar acoustic scenario, however, only with a $\theta_{error}=10.0^\circ$.

\clearpage

As it can be seen, some runs with the previous optimization mechanism presented in \cite{rascon2026multi} were able to reach the $[\theta_{correct}]$ range, but only after a considerably longer time than the optimization mechanism proposed in this work. Additionally, the proposed optimization mechanism provides considerably less variation between runs, which results in an optimization behavior that is considerably more stable than its previous counterpart. All of this while requiring much less complexity to carry out the optimization process, which can be clearly seen when comparing Algorithms \ref{alg:optimization2} and \ref{alg:newoptimization}.

\subsection{Performance in Different Acoustic Settings}

To evaluate the robustness of the $[\theta_{range}=20.0,\theta_{steps}=5]$ configuration across different acoustic settings, three other recordings from ``Office A'' scenario were selected with varying amounts of sources present, including the target speech source (2, 3, and 4). Additionally, three other recordings, with varying amounts of sources present (2, 3, and 4), were chosen from the ``Department Store'' scenario to evaluate the aforementioned optimization configuration. To facilitate the calculation of the evaluation metrics, all these recordings were chosen such that the target speech source still is located at $0^\circ$. However, it is important to state that in the case of the recordings with 3 sources present, the location of the target speech source was set to $4^\circ$ to compensate for a possible mistake in the true location reported in the documentation of the AIRA corpus, which admittedly was measured by hand and is prone to human error \cite{rascon2018acoustic}.

The reader is reminded that the ``Office A'' scenario has a low amount of noise (52 dB SPL) with a moderately high level of reverberation ($\tau_{60}=0.20s$), where as the ``Department Store'' scenario has a high level of noise (63 dB SPL) with a moderate level of reverberation ($\tau_{60}=0.16s$).

\begin{table}[h]
	\caption{Evaluation of $[\theta_{range}=20.0,\theta_{steps}=5]$ configuration in different acoustic scenarios.}\label{tab:diffscenarios}
	\begin{tabular*}{\textwidth}{@{\extracolsep\fill}cccccccc}
		\toprule%
		& & \multicolumn{3}{@{}c@{}}{Department Store} & \multicolumn{3}{@{}c@{}}{Office A} \\
		\cmidrule{3-5}\cmidrule{6-8}%
		\# sources & $\theta_{error}$ & RMS $\downarrow$ & $\sigma_{RMS}$ $\downarrow$ & $t_{opt}$ $\downarrow$ & RMS $\downarrow$ & $\sigma_{RMS}$ $\downarrow$ & $t_{opt}$ $\downarrow$ \\
		\midrule
		  &  5$^\circ$ & 2.37$^\circ$ & 0.13$^\circ$ & -        & 2.83$^\circ$ & 0.23$^\circ$ & -       \\
		2 & 15$^\circ$ & 2.42$^\circ$ & 0.18$^\circ$ &  6.97 s. & 3.00$^\circ$ & 0.23$^\circ$ & 3.12 s. \\
		  & 25$^\circ$ & 2.33$^\circ$ & 0.08$^\circ$ & 10.56 s. & 2.88$^\circ$ & 0.17$^\circ$ & 9.11 s. \\
		\midrule
		  &  5$^\circ$ & 3.76$^\circ$ & 0.17$^\circ$ & -        & 3.13$^\circ$ & 0.26$^\circ$ & -        \\
		3 & 15$^\circ$ & 3.54$^\circ$ & 0.23$^\circ$ & 10.76 s. & 3.36$^\circ$ & 0.41$^\circ$ &  7.29 s. \\
		  & 25$^\circ$ & 3.67$^\circ$ & 0.34$^\circ$ & 11.64 s. & 3.31$^\circ$ & 0.33$^\circ$ & 11.65 s. \\
		\midrule
		  &  5$^\circ$ & 4.85$^\circ$ & 0.20$^\circ$ & -        & 4.31$^\circ$ & 0.22$^\circ$ & -        \\
		4 & 15$^\circ$ & 4.87$^\circ$ & 0.08$^\circ$ &  9.10 s. & 4.49$^\circ$ & 0.44$^\circ$ & 10.81 s. \\
		  & 25$^\circ$ & 4.84$^\circ$ & 0.17$^\circ$ & 18.01 s. & 4.64$^\circ$ & 0.35$^\circ$ & 14.68 s. \\
		\botrule
	\end{tabular*}
\end{table}

As it can be seen, when increasing the amount of sources the accuracy falls, which is to be expected since MVDR faces an increased amount of cumulative interference energy when the number of sources increases. Thus, the quality estimations provided by Squim are expected to have more errors in such cases. However, it seems that the accuracy falls by only $\approx 1^\circ$ for each additional source, which is as good trade-off considering that what is gained is that varying $\theta_{error}$ with the same number of sources does not alter the accuracy (RMS). This is very important since it shows that the $[\theta_{range}=20.0,\theta_{steps}=5]$ configuration is robust against localization errors, which is the prime concern in the application areas that multi-agent auditory scene analysis can benefit.

Moreover, in all circumstances, the RMS is below $5^\circ$, which establishes that the $[\theta_{range}=20.0,\theta_{steps}=5]$ configuration reaches and stays in the $[\theta_{correct}]$ range in a consistent manner regardless of acoustic scenario, $\theta_{error}$, or number of sources present.

Additionally, the optimization time ($t_{opt}$) increases as $\theta_{error}$ is increased, which is to be expected since the starting point is farther away from the $[\theta_{correct}]$ range. However, in all scenarios (except the for most difficult scenario, with 4 sources and $\theta_{error}=25^\circ$) the optimization time is below $15.0$ s., which is a considerable improvement over the optimization mechanism presented in \cite{rascon2026multi}, as shown in Figure \ref{fig:optimorig}.

Moreover, having the optimization time below $15.0$ s. is equivalent to one of the restrictions set in the characterization stage of this evaluation. Along with the fact that in all circumstances the variance ($\sigma_{RMS}$) is comparable to the value reported in Table \ref{tab:reccomb}, it could be then deduced that the $[\theta_{range}=20.0,\theta_{steps}=5]$ configuration provides a comparable level of speed and stability as when it was being characterized using the first recording of the ``Office A'' scenario, regardless of acoustic scenario, $\theta_{error}$, or number of sources present. This implies that the optimization process is expected to `behave' similarly in different acoustic scenarios.

\section{Conclusion}\label{sec:conclusion}

A real-time auditory scene analyzer (ASA) carries out in conjunction the tasks of locating, separating, and classifying sound sources in the environment, with important benefits to application areas such as bioacoustics, hearing aid research, search and rescue scenarios, human-robot interaction, etc. Recently, there have been efforts to model an ASA as a multi-agent system, where agents (each in charge of a particular task) interact with one another to provide local error robustness at the global level, while reducing overall complexity. In these efforts, the feedback loop between the localization and separation agents was shown to be able to correct the localization error by optimizing the separation quality using the reference-free quality estimation model called Squim. However, it was also shown to be very slow to reach the correct location, while being relatively complex.

In this work, a new optimization mechanism is proposed that exploits the interaction between the quality estimation agent and the optimization agent (which are part of the aforementioned localization-separation feedback loop). The quality estimation agent, instead of carrying out one quality estimation steered a given location at each optimization iteration, it estimates several quality estimations, each steered at a different location (spread from a given middle location). The optimization agent is reduced to a weighted average, using the quality estimated values as weights, carrying out a type of simplified interpolation inside a given location range. Although there is not an optimization process instructed anywhere in the two agents, an optimization behavior emerges from their interaction, with the possibility of being `calibrated' by ways of modifying the size of the tested location range ($\theta_{range}$) and how many quality estimations are carried out inside it ($\theta_{steps}$). 

The evaluation showed that the proposed optimization mechanism provided good results in different real-life acoustic scenarios, with varying noise and reverberation levels, and with varying amounts of sound sources. More importantly, it showed to be more robust to larger localization errors, provides stabler results, and be faster to reach the correct location than the previous optimization mechanisms, by a wide margin in all accounts, while being considerably less complex.

It is important to mention that, given that multiple quality estimations ($\theta_{steps}$) are being carried out, in the proposed optimization mechanism the quality estimation agent has a considerably higher response time than the previous optimization mechanisms. However, it is still able to satisfy the requirement for real-time operation. Additionally, given the parallel nature of multi-agent systems, the separation latency (the time it takes to process a given audio window) is not affected. The quality estimation agent response time does affect the optimization time (the time the whole system takes to reach the correct location), but an in-depth characterization of the $[\theta_{range},\theta_{steps}]$ optimization parameters resulted in a set of values that provided a good balance between optimization time, quality estimation response time, and localization accuracy and stability: $[\theta_{range}=20.0,\theta_{steps}=5]$.

There are two factors that impact the quality estimation response time which, in turn, impacts the optimization time. The first factor is the amount of quality estimations carried out at every optimization iteration ($\theta_{steps}$), which has already been characterized in this work. The second factor is the response time of the reference-free quality estimation model, Squim. It is important to reiterate that, although it is able to run in real-time, such a model does not consider causality internally; it assumes that it is being fed a complete recording. Thus, as it was shown in \cite{rascon2026multi}, it requires a considerable amount of data to provide even moderately stable results, with $3.0$ s. being a good balance point between stability and latency. For future work, other reference-free quality estimation models will be explored to find one that provides smaller latencies and stabler results. Another alternative is to modify the Squim model to consider causality which may reduce its latency, while keeping its current level of stability.

\backmatter

\section*{Declarations}

\bmhead{Competing Interests}

The author declares that they have no known competing financial interests or personal relationships that could have appeared to influence the work reported in this paper.

\bmhead{Funding}

This work was supported by PAPIIT-UNAM through the grant IN100624.

\bmhead{Data/Code availability}

The code of all the agents of the proposed auditory scene analyzer can be found in \href{https://github.com/balkce/masa}{https://github.com/balkce/masa}.

\bibliography{references}% common bib file

\end{document}